\documentclass[10pt]{IEEEtran}
\usepackage{cite}
\usepackage{amsmath,amssymb}
\usepackage{algorithmic}
\usepackage{graphicx,color}
\usepackage{textcomp}
\usepackage{hyperref}
\usepackage{mathtools}
\usepackage{multirow}
\def\BibTeX{{\rm B\kern-.05em{\sc i\kern-.025em b}\kern-.08em
    T\kern-.1667em\lower.7ex\hbox{E}\kern-.125emX}}
\AtBeginDocument{\definecolor{green}{cmyk}{1,0,1,0.5}}

\begin{document}

\title{A Thermal-Electrical Co-Optimization Framework for Active Distribution Grids with Electric Vehicles and Heat Pumps
\thanks{This project has received funding from the EU’s Horizon Europe Framework Programme (HORIZON) under the GA n. 101120278 - DENSE.}
}
\author{\IEEEauthorblockN{Savvas Panagi\IEEEauthorrefmark{1}\IEEEauthorrefmark{2},  
Chrysovalantis Spanias\IEEEauthorrefmark{1},
and
Petros Aristidou\IEEEauthorrefmark{2}}\\ 
\IEEEauthorblockA{
\IEEEauthorrefmark{1}Distribution System Operator, Electricity Authority of Cyprus, Nicosia, Cyprus.\\
\IEEEauthorrefmark{2}Dept. of Electrical \& Computer Engineering, \& Informatics, Cyprus University of Technology, Limassol, Cyprus\\
Emails: cspanias@eac.com.cy, 
\{savvas.panagi, petros.aristidou\}@cut.ac.cy}
}

\maketitle

\begin{abstract}
The growing electrification of transportation and heating through Electric Vehicles (EVs) and Heat Pumps (HPs) introduces both flexibility and complexity to Active Distribution Networks (ADNs). These resources provide substantial operational flexibility but also create tightly coupled thermal–electrical dynamics that challenge conventional network management. This paper proposes a unified co-optimization framework that integrates a calibrated 3R2C grey-box building thermal model into a network-constrained Optimal Power Flow (OPF). The framework jointly optimizes EVs, HPs, and photovoltaic systems while explicitly enforcing thermal comfort, Distributed Energy Resource (DER) limits, and full power flow physics. To maintain computational tractability, Second-Order Cone Programming (SOCP) relaxations are evaluated on a realistic low-voltage feeder. The analysis shows that, despite network heterogeneity violating some theoretical exactness conditions, the relaxation remains exact in practice. Comparative assessments of convex DistFlow, bus injection, and branch flow formulations reveal that convex DistFlow achieves sub-second runtimes and near-optimal performance even at high DER penetration levels. Simulations confirm the effectiveness of coordinated scheduling, yielding reductions of 41\% in transformer aging, 54\% in losses, and complete elimination of voltage violations, demonstrating the value of integrated thermal-electrical coordination in future smart grids.
\end{abstract}


\maketitle

\section{INTRODUCTION}
\label{sec:introduction}

\IEEEPARstart{T}{he} ongoing transition toward electrified residential energy systems, driven by the integration of Electric Vehicles (EVs) and Heat Pumps (HPs), presents both challenges and opportunities for distribution system operators \cite{7155600}. These assets offer flexibility for balancing and voltage regulation \cite{masuta2012supplementary}, but they also introduce variability and coupling across the thermal, transportation, and electrical domains. Efficient management, therefore, requires accurate modeling and scalable optimization techniques that capture the behavior of Distributed Energy Resources (DERs) and the nonlinear nature of power flows.

The literature has proposed a variety of methods to coordinate flexibility assets such as EVs, HPs, and photovoltaic (PV) systems to enhance network operation or reduce costs. In \cite{petrucci2025coordinated}, a nonlinear model predictive control is applied to co-optimize residential EV and HP operation, but it lacks grid awareness and does not account for distribution network constraints. The approach in \cite{hao2024optimal} addresses this by incorporating electro-thermal modeling and full network representation into a comprehensive scheduling framework; however, it relies on Particle Swarm Optimization, a heuristic method without global optimality guarantees and with high computational demands for large-scale or real-time use. Studies such as \cite{muhssin2018dynamic,georges2017residential} highlight the potential of residential HPs for services like frequency response and market-driven flexibility, yet they omit system-level scheduling and network-constrained optimization. Moreover, current control strategies often depend on water tank flexibility, increased self-consumption from Renewable Energy Sources (RES), or overly simplified user comfort models \cite{bhattarai2014demand,kemmler2020design,beck2017optimal}, but these are seldom practical due to high upfront costs, limited modeling accuracy, or their confinement to local rather than centralized schemes.

Similar developments have occurred in the domain of EVs, which are another source of demand-side flexibility. Various strategies manage EV charging through unidirectional or bidirectional flows, enabling services such as Vehicle-to-Grid (V2G) \cite{TAN2016720,milad_falahi,5649138}, voltage regulation \cite{kotsonias2025operational}, frequency support \cite{KUSHWAHA2023106944}, and load balancing \cite{Sam}. Aggregators are frequently employed to coordinate EV fleets, streamlining their participation in energy markets \cite{GONZALEZVENEGAS2021111060}. Collectively, these capabilities make EVs a cost-effective tool for enhancing grid flexibility \cite{habib2018assessment}.

However, the effective integration of such flexible resources relies on accurate modeling of the distribution network. The power flow problem is inherently nonlinear, non-convex, and NP-hard \cite{bienstock2019strong}, posing significant challenges for optimization due to non-global solutions, long runtimes, and possible solver failure. To address this, several convex relaxations of the Alternating Current (AC) power flow equations have been proposed \cite{farivar2013branch,kocuk2016strong}.

\subsection{GAP ANALYSIS}

Current research approaches exhibit limitations in three fundamental areas that constrain the practical implementation of flexible energy systems focusing on EVs, HPs, and PVs, which constitute the dominant and most controllable residential assets in the distribution level \cite{venegas2021active}. First, the thermal-electrical interface typically lacks adequate integration: existing studies generally do not couple HP operation with rigorous building thermodynamics, resulting in oversimplified models that cannot fully capture real-world performance dynamics. This disconnect limits the effectiveness of current optimization strategies for actual deployment scenarios. 

Second, the fragmented treatment of DERs constrains the potential for system-wide optimization: most Active Distribution Network (ADN) scheduling frameworks consider only subsets of available flexibility assets, limiting the realization of multiplicative benefits from coordinated multi-asset operation and leaving substantial economic and operational value underutilized. 
Finally, the computational demands of existing network-aware approaches pose scalability challenges: the predominant reliance on heuristic methods and nonlinear solvers foregoes global optimality guarantees and can render them computationally intensive for real-time operation at distribution scale, thereby constraining their transition from academic studies to practical grid management tools.

\subsection{PAPER CONTRIBUTIONS}

This work addresses the identified gaps through a novel modeling and optimization framework that explicitly couples building thermal dynamics with HP electrical operation, coordinates all major distribution-level flexibility assets, and applies a convex power flow formulation. A high-fidelity calibrated 3R2C grey-box model is developed that represents building thermodynamics and maps thermal behavior to HP electrical consumption, enabling precise thermal-electrical optimization. This thermal model is embedded within a comprehensive scheduling framework that simultaneously coordinates HPs, EVs, and PVs under network constraints by relaxing the DistFlow equations to an SOCP. 

Therefore, the main contributions of this work are threefold:
\begin{itemize}
\item \textit{Thermal-electrical coupling:} A unified framework that integrates calibrated 3R2C grey-box models directly into electrical optimization, enabling precise mapping from thermal comfort constraints to HP electrical flexibility.
\item \textit{Comprehensive multi-asset coordination:} Simultaneous scheduling of HPs, EVs, and PVs within a single network-constrained Optimal Power Flow (OPF) framework, capturing realistic operational constraints and inter-asset dependencies.
\item \textit{Scalable convex power flow formulation:} Computational assessment and exactness analysis of SOCP relaxation for realistic LV distribution networks under varying DER penetration levels, demonstrating substantial computational advantages over nonlinear formulations.
\end{itemize}

The remainder of the paper is structured as follows. Section~\ref{sec:model_framework} presents the modeling framework, including DER constraints, the 3R2C thermal model for HPs, and the DistFlow power flow formulation with its SOCP relaxation. The case study setup, including network characteristics, building thermodynamics, and DER data, is described in Section~\ref{sec:case_study}. Section~\ref{sec:Results_Discussions} presents the computational comparison of OPF formulations, exactness analysis, and network operation improvements. Finally, Section~\ref{sec:conclusions} summarizes the key findings and outlines future research directions.

\section{MODELING FRAMEWORK}
\label{sec:model_framework}

\subsection{NOTATIONS}

Let \( \mathbb{C} \) denote the set of complex numbers. Consider a directed graph \( G = (\mathcal{N}, \mathcal{L}) \), where the set of buses is \( \mathcal{N} = \{0, 1, \dots, n\} \) with \( |\mathcal{N}| = n + 1 \), and the set of lines is \( \mathcal{L} \), with \( |\mathcal{L}| = m \). A line directed from bus \( i \) to \( j \) is denoted \( i \sim j \in \mathcal{L} \) and is modeled using the series circuit with admittance: \( y^s_{ij} = y^s_{ji} \in \mathbb{C} \). The complex power flow \( S_{ij} \) and current \( I_{ij} \) represent the power and current from bus \( i \) to \( j \), \( \forall i \sim j \in \mathcal{L} \). Although graph orientation is arbitrary, it is typically chosen so that bus $0$ is the root node, with all lines directed outward to form a directed tree. The set of transformers is defined as $\mathcal{R}$, while DERs are defined as $\mathcal{D}$. 
The line impedance is $z_{ij} = z_{ji} \coloneqq 1/y^s_{ij}$, and the complex power injection at bus $j$ is $s_j \coloneqq p_j + \mathbf{i}q_j$, with voltage phasor $\mathrm{V_j}$. $\ell_{ij} \coloneqq |I_{ij}|^2$ and $v_j \coloneqq |V_j|^2$ define current and voltage magnitude square. $\overline{\bullet}$ $/$ $\underline{\bullet}$ $/$ $\widetilde{\bullet}$, defined upper/lower bound and estimated value of a quantity, respectively. 
The set $\Omega^{\mathcal{D}}_\theta$ denotes the subset $\Omega^{\mathcal{\theta}} \subseteq \Omega^{\mathcal{D}}$, 
for all $\theta \in \Theta$, where 
$\Theta = \{\text{RES}, \text{EVs}, \text{HPs}\}$ 
represents the distributed resource types. Finally, $H$ denotes the time step.

\subsection{DISTRIBUTED ENERGY RESOURCES}

We consider a set of electricity-consuming and producing resources. 
Each resource is classified into a specific group within the type set $\Omega^{\mathcal{D}}_\theta, \forall \theta \in \Theta$. Each resource has defined upper and lower limits for active power as described in~\eqref{eq:active_power_limits}. 
RES units have an additional constraint on their active power output due to exogenous factors such as solar irradiance and wind speed, making their maximum available power time-dependent as described in~\eqref{eq:active_power_res_limits}. 
Resources may operate at any level within these limits without ramp-rate constraints, as the selected time step is sufficiently large for DERs to adjust their operating points.

Reactive power control, based on the IEEE~1547--2018 standard~\cite{8332112}, supports several local operating modes:  
(i) constant Power Factor (PF),  
(ii) voltage--reactive power (volt--var),  
(iii) active power--reactive power (watt--var), and  
(iv) constant reactive power.  
These modes are expressed mathematically in~\eqref{eq:reactive_power_function}, where $Q$ is the reactive power function and $x$ is the state variable, e.g., for constant PF:  
$ Q = P \cdot \tan(\cos^{-1}(PF)),~x = P. $
Alternatively, reactive power can be set via a central controller. 
In all cases, operation must comply with nominal limits~\eqref{eq:reactive_power_constraints}. 
In this work, central reactive control is applied to PVs and EVs in either inductive or capacitive mode, while a constant inductive PF is assumed for HPs.

\begin{align}
\label{eq:active_power_limits}
&\underline{P}_{der} \leq P_{der,t} \leq \overline{P}_{der}, \quad && \forall der \in \Omega^{\mathcal{D}}, t \in \Omega^\mathcal{T} \\
& \underline{Q}_{der} \leq Q_{der,t} \leq \overline{Q}_{der}, \quad && \forall der \in \Omega^{\mathcal{D}}, t \in \Omega^\mathcal{T} 
\label{eq:reactive_power_constraints} \\
\label{eq:active_power_res_limits}
& P_{der,t} \leq \widetilde{P}_{der,t}, \quad && \forall  der \in  \Omega^{\mathcal{D}}_{EV}, t \in \Omega^\mathcal{T} \\
\label{eq:reactive_power_function}
& Q_{der,t} = {Q(x)}, \quad && \forall der \in \Omega^{\mathcal{D}}, t \in \Omega^\mathcal{T} 
\end{align}

\subsection{ELECTRIC VEHICLES}

For EVs, additional constraints apply to the charging process, which depends on user preferences and connection-related information, such as arrival and departure times ($\widetilde{t}^{arr}_{ev}$, $\widetilde{t}^{dep}_{ev}$). 
The state-of-charge (SoC) transition function for an EV follows the equation in~\eqref{eq:soc_transition}, where the energy level at time $t$ is determined by its previous state and the charging power over the time step $H$. 
The EV's stored energy is subject to upper and lower limits, as expressed in~\eqref{eq:soc_ev_limits}, which depend on both user-defined constraints and manufacturer-imposed technical limits, with the most restrictive condition applied to ensure efficient operation. 
Equations~\eqref{eq:power_completeness_1} and~\eqref{eq:power_completeness_2} enforce completeness by stating that there can be no active or reactive power flow when the EV is not connected. 
This allows certain variables to be determined directly, removing them from the optimization.

The following constraints are added $\forall ev \in \Omega^{\mathcal{D}}_{EV}$, $t \in \Omega^\mathcal{T}$:
\begin{align}
\label{eq:soc_transition}
&E_{ev,t} = E_{ev,t-1} + x_{ev,t} \cdot P_{ev,t}\cdot H \\
\label{eq:soc_ev_limits}
&\underline{E}_{ev} \leq E_{ev,t} \leq \overline{E}_{ev} \\
\label{eq:power_completeness_1}
& P_{ev,t},Q_{ev,t}, x_{ev,t} = 0  \quad \text{if } t \notin [\tilde{t}_{ev}^{arr}, \tilde{t}_{ev}^{dep}] \\
\label{eq:power_completeness_2}
& x_{ev,t}=1 \quad \text{if } t \in [\tilde{t}_{ev}^{arr}, \tilde{t}_{ev}^{dep}]
\end{align}
where $x_{ev,t}$ is a binary parameter that equals 1 when the EV is online, and 0 otherwise.

\subsection{HEAT PUMPS}

\begin{figure}
    \centering
    \includegraphics[width=\linewidth]{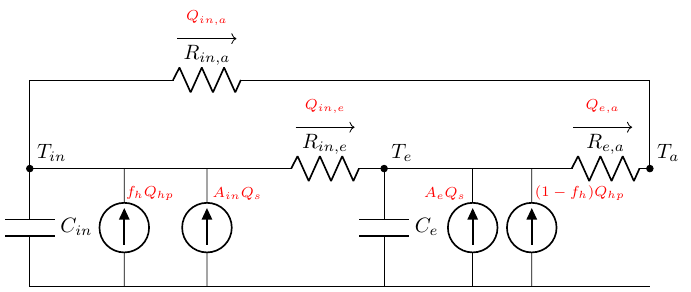}
    \caption{Thermal network representation of the 3R2C grey-box model.}
    \label{fig:3r2c_model}
\end{figure}

The operation of HPs is intrinsically linked to the thermal dynamics of the buildings they serve. To capture this bidirectional coupling, a grey-box Resistance–Capacitance (RC) thermal model is employed, based on a 3R2C configuration \cite{harb2016development} that represents both indoor air and building envelope dynamics, as illustrated in Fig.~\ref{fig:3r2c_model} and described by~\eqref{eq:Q_hp,t} and~\eqref{eq:T_e,t}. This model describes the evolution of indoor temperature as a function of thermal capacity, resistance parameters, and external influences such as ambient temperature and solar irradiance. In addition to modeling HP operation, the formulation includes an explicit thermal-to-electrical mapping through~\eqref{eq:thermal_electrical_map}, which converts heating demand into electrical power consumption based on the operating coefficient of performance. To ensure compliance with user-defined comfort preferences, a constraint on indoor temperature is imposed, as described in~\eqref{eq:Tin_limits}.

The following constraints are added $\forall hp \in \Omega^{\mathcal{D}}_{HP}$, $t \in \Omega^\mathcal{T}$:

\begin{align} 
    \label{eq:thermal_electrical_map}
    {P}_{hp,t} & = \frac{{Q}_{hp,t}}{COP_{hp,t}(T_{e,t})}\\
    \label{eq:Q_hp,t}
    {Q}_{hp,t} &= \frac{T_{in,t+1}-T_{in,t}}{H} \cdot \frac{C_{in}}{f_h} - \frac{T_{e,t}-T_{in,t}}{R_{in,e} \cdot f_h} - \\ \nonumber &\quad \frac{T_{a,t}-T_{in,t}}{R_{in,a} \cdot f_h} - \frac{A_{in} \cdot Q_{s,t}}{f_h} \\
    \label{eq:T_e,t}
    T_{e,t+1} & = T_{e,t} + H \cdot \bigg( 
     \frac{T_{in,t} - T_{e,t}}{C_e \cdot R_{in,e}}  + \frac{T_{a,t} - T_{e,t}}{C_e \cdot R_{in,e+}} \\ \nonumber
    &  + \frac{(1-f_h) \cdot Q_{hp,t}}{C_e} + \frac{A_e \cdot Q_{s,t}}{C_e}  
\bigg) \\
    \label{eq:Tin_limits}
    T_{in,t}^{min} & \leq T_{in,t} \leq T_{in,t}^{max} \quad [^{\circ} C] 
\end{align}
where $T_{in,t}$ [$^{\circ} C$] denotes the indoor air temperature, $T_{e,t}$ [$^{\circ} C$] is the temperature of the building envelope, and $T_{a,t}$ [$^{\circ} C$] is the ambient outdoor temperature. The term $Q_{s,t}$ [W/m\textsuperscript{2}] represents the horizontal solar irradiance, and $Q_{hp,t}$ [W] is the total heating demand, distributed between the internal and external thermal nodes according to the splitting factor $f_h \in [0,1]$. The parameters $C_{in}$ [J/K] and $C_e$ [J/K] denote the thermal capacitance of the indoor air and the envelope, respectively. Thermal resistances between the indoor zone and the envelope and ambient air are given by $R_{in,e}$ [K/W] and $R_{in,a}$ [K/W], respectively, and $R_{e,a}$ [K/W] denotes the thermal resistance between the envelope and ambient air. $A_{in}$ and $A_e$ [m\textsuperscript{2}] define the effective surface areas for solar gains acting on the indoor and envelope nodes. The simulation timestep $H$ [s] is used to discretize the thermal dynamics. Model parameters for each representative building are estimated through a data calibration algorithm, that used least squares errors, see~\cite{tunable_greybox_thermal_preprint} for more details.

\subsection{NETWORK AND POWER FLOW CONSTRAINTS}
\label{sec:distflow}

The Branch Flow Model (BFM) equations are adopted, with the shunt elements of the $\Pi$ line model assumed to be zero, and a radial network topology considered. After an exact reformulation, voltage and current angles are eliminated from the formulation. This power flow formulation, known as the \textit{DistFlow} model, is given in \eqref{eq:distflow_p}–\eqref{eq:distflow_l} and was first proposed in~\cite{baran1989optimal}. Voltage angles can be recovered from the original equations, but this is valid \textit{only} for radial networks, as proven in~\cite{farivar2013branch}.
The voltage magnitude limits in~\eqref{eq:voltage_limits} ensure that the voltage at each bus $i$ remains within safe bounds, preserving power quality and protecting connected equipment. The ampacity limits in~\eqref{eq:line_current_limits} ensure that the current magnitude on each line $(i,j)$ does not exceed the thermal capacity $\overline{I}_{ij}$, thereby preventing overheating and damage. The line model simplification may cause discrepancies in the accuracy of $I_{ij,t}$; further details on this inconsistency are provided in~\cite{christakou2017ac}.

The following network constraints are added  $\forall t \in \Omega^\mathcal{T}$:

\begin{align}
\label{eq:distflow_p}
p_j &= \sum_{k: j \sim k} P_{jk} - \sum_{i: i \sim j} \left( P_{ij} - r_{ij} \ell_{ij} \right) + g_j v_j, && \forall j \in \mathcal{N} 
\end{align}
\begin{align}
\label{eq:distflow_q}
q_j &= \sum_{k: j \sim k} Q_{jk} - \sum_{i: i \sim j} \left( Q_{ij} - x_{ij} \ell_{ij} \right) + b_j v_j, && \forall j \in \mathcal{N} \\
\label{eq:distflow_v}
v_j &= v_i - 2 (r_{ij} P_{ij} + x_{ij} Q_{ij}) + (r_{ij}^2 + x_{ij}^2) \ell_{ij}, && \forall (i,j) \in \mathcal{L} 
\end{align}
\begin{align}
\label{eq:distflow_l}
\ell_{ij} &= \frac{P_{ij}^2 + Q_{ij}^2}{v_i}, && \forall (i,j) \in \mathcal{L}\\
\label{eq:voltage_limits}
& \underline{V}_j^2  \leq |V_{j,t}|^2, \quad &&\forall j \in \mathcal{N} \\
\label{eq:line_current_limits}
&|I_{ij,t}|^2 \leq \overline{I}_{ij}^2, \quad &&\forall (i,j) \in \mathcal{L}
\end{align}

\subsection{CONIC RELAXATION OF DISTFLOW}
\label{sec:conic_relaxation}
This relaxation is obtained by applying SOCP to the non-convex ~\eqref{eq:distflow_l}. Further details on this relaxation are provided in \cite{farivar2013branch}. The proof of exactness for radial networks assumes either no upper bounds on load consumption at each bus or no lower bounds on power injection, both conditions being equivalent in this context. These assumptions generally do not hold in realistic settings, especially in systems with static loads or limited demand-response capacity. However, subsequent studies have shown that the same SOCP relaxation can remain exact under alternative sufficient conditions better aligned with practical networks. For instance,~\cite{li2012exact} presents provable conditions based on cumulative line resistance/reactance ratios and nodal voltage levels, which can be evaluated \textit{a priori}. In particular, it provides a simple, though conservative, voltage-based criterion that guarantees exactness even in the absence of upper voltage bounds. The case study analysis will illustrate practical aspects of these conditions, and the reader is referred to~\cite{li2012exact} for a comprehensive theoretical treatment.

\subsection{OBJECTIVE FUNCTION}

The proposed AC OPF formulation schedules EVs, HPs, and PV units to minimize line losses ($C_{llosses}$) and transformer losses ($C_{tlosses}$), while also minimize photovoltaic curtailment ($C_{der}$). A weighting coefficient $\alpha$ is applied to the curtailment cost term, set sufficiently large to prioritize curtailment reduction over power loss minimization.

The following objective function are added $\forall t \in \Omega^\mathcal{T}$:
\begin{equation}
\begin{aligned}
& \min_{P_{der}, Q_{der}, l, v}  \alpha \cdot C_{der} + C_{llosses} + C_{tlosses} \\
&s.t.: \\
& C_{der} = \sum_{der \in \Omega^{\mathcal{D}}_{RES}}  (\widetilde{P}_{der,t} - P_{der,t}) \\
& C_{llosses} = \sum_{(i,j) \in \mathcal{L}} r_{ij} \ell_{ij,t}, \quad C_{tlosses} = \sum_{(i,j) \in \mathcal{R}} P_{fe} +r_{ij} \ell_{ij,t} 
\end{aligned}
\label{eq:ROPF}
\end{equation}

\section{CASE STUDY}
\label{sec:case_study}

\subsection{TRANSFORMER AGING METRIC}
\label{sec:transformer_aging}

The aging process of transformers is influenced by thermal effects arising from continuous loading. The transformer's lifespan is closely linked to the winding's hottest-spot temperature, which drives insulation system degradation. The IEEE C.57.91-2011~\cite{6166928} standard models this relationship using an exponential expression for the aging acceleration factor. In this work, we focus on degradation effects under operating conditions where transformer loading remains below its nominal rated power. Voltage and frequency influences should also be considered when defining operational limits beyond the nameplate rating (see IEEE Std C57.12.00).

The relationship between insulation aging and temperature follows an Arrhenius-based model. The aging acceleration factor $F_{AA}$ represents the relative insulation degradation rate compared to nominal aging at a reference hotspot temperature of 110$^{\circ} C$, as defined in~\eqref{eq:aging_acceleration_factor}. A value $F_{AA} > 1$ indicates faster aging than nominal (above 110$^{\circ} C$), while $F_{AA} < 1$ implies slower aging (below $^{\circ} C$). For the scheduling horizon, the equivalent aging factor~\eqref{eq:equiv_aging_factor} is used to evaluate the impact over the entire simulation period. The $\theta_H$ variable is calculated following the method described in Section~7.2 of~\cite{6166928}.

\begin{align}
\label{eq:aging_acceleration_factor}
&F_{AA,t} = \exp\left( \frac{15000}{383} - \frac{15000}{\theta_{H,t} + 273} \right) \\ 
\label{eq:equiv_aging_factor}
&F_{EQA} = \left ( \sum\limits_{t=1}^{\mathcal{T}} F_{AA,t} \cdot H_t \right)  / \sum\limits_{t=1}^{\mathcal{T}} H_t 
\end{align}
where:
\begin{itemize}
  \item $F_{AA}$: aging acceleration factor (dimensionless),
  \item $\theta_H$: winding hottest-spot temperature [$^{\circ} C$].
\end{itemize}

\subsection{HOUSEHOLD LOAD AND PV GENERATION PROFILES}

In this study, electricity consumption data from 68 residential customers in Cyprus, recorded every 30-minute over a year, are used. To focus on general consumption behavior and exclude households likely using electric heating, a seasonal filtering method compares each customer's total load consumption in winter to that in autumn. Customers with a winter-to-autumn ratio above 1.6 are excluded, being assumed to operate heating systems. After filtering, 57 customers remain, forming a refined dataset of non-heating customers. This subset is used to randomly assign Basic Load Profiles (BLPs) to feeder customers. January is selected for analysis as it has the highest heating demand and the most stressful winter BLP.

Statistical data are used to randomly distribute installed PV capacity across residential rooftops according to the probability mass function provided by the Cyprus Distribution System Operator (DSO). The maximum capacities for single-phase and three-phase systems are 4.16~kWp and 10.4~kWp, respectively, based on DSO limits. Due to the geographic proximity of households, the PV generation profile is assumed identical but scaled to match the installed capacity. Based on DSO data, in suburban areas, 70\% of residential buildings are detached houses with a three-phase supply, and the rest have a single-phase supply. For the uncontrollable operation, a constant inductive $ PF = 0.95 $ is assumed.

\subsection{EV MODELING AND DATA}
\label{sub:EV_modeling}

The driving patterns, including departure and arrival times and trip distances, were collected through a survey targeting car owners who commute daily in Cyprus \cite{sokr_pissa, panagi2025optimal}. The initial SoC values for EVs are randomly and uniformly assigned within predefined minimum and maximum limits of 20\% and 80\%, respectively. The specific EV type plays a key role in converting driving patterns into charging profiles, as it determines battery capacity and energy consumption rate. For this study, average technical specifications derived from European EV sales data are used, assuming an average consumption of 17~kWh/100km and a driving range of up to 400~km.
For households with three-phase connections, 11~kW chargers are assumed, while single-phase connections use 3.7~kW chargers. By default, 50\% of residential customers are assumed to own and operate EVs, unless stated otherwise. The minimum SoC is set to 30\% and the maximum to 80\%; for long trips, this range is extended to 20\% and 90\%, respectively. Further details on the dataset can be found in~\cite{panagi2025impact}.

\subsection{HEAT SYSTEMS AND BUILDING THERMODYNAMICS}

\begin{figure}
    \centering
    \includegraphics[width=1\linewidth]{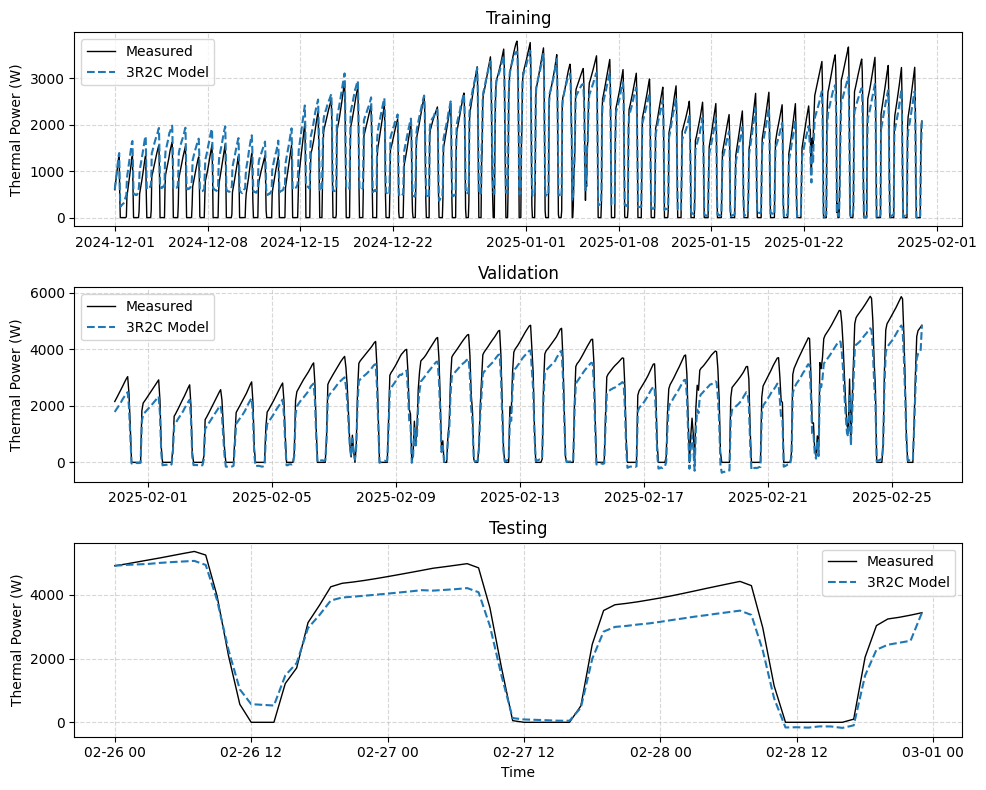}
    \caption{RC model tuning performance using the identified 3R2C grey-box model across three phases: training, validation, and testing.}
    \label{fig:training_3r2c}
\end{figure}

This study assumes an Air-Source Heat Pump (ASHP) for all buildings, justified by Cyprus's mild Mediterranean climate and the lower economic viability of ground-source HPs in such conditions~\cite{gaur2021heat}. Three building types are considered: Single-Family House (SFH), OFFice (OFF), and Trade/Commercial Building (TRA). It is assumed that 50\% of total load customers have an HP system. Of these, 60\% are SFHs, 20\% are commercial buildings, and the remainder are offices. The tuning procedure relies on a least-squares error metric to calibrate the model parameters; a detailed description of the tuning methodology is provided in~\cite{tunable_greybox_thermal_preprint}. The selected model tuning parameters are listed in Table~\ref{tab:3r2c_results}, and the training, validation, and testing performance of the SFH model is shown in Fig.~\ref{fig:training_3r2c}. Additional building characteristics are available in the TABULA~\cite{TABULA_Project} and Hotmaps~\cite{hotmaps_project} projects. Synthetic learning data were generated using parameter modeling proposed by the Ambience project~\cite{jankovic2021database}.

\begin{table}[htbp]
\centering
\caption{3R2C Model Parameters and Root Mean Square Error (RMSE) for Three Building Types}
\label{tab:3r2c_results}
\begin{tabular}{lccc}
\hline
\textbf{Parameter / Metric} & \textbf{SFH} & \textbf{OFF} & \textbf{TRA} \\
\hline
$R_{in,e} [K/W]$   & 0.00038 & 0.00170 & 0.00028 \\
$R_{in,a} [K/W]$   & 0.04996 & 0.03956 & 0.04785\\
$R_{e,a} [K/W]$    & 0.00307 & 0.00118 & 0.00153\\
$C_{in} [J/K]$     & $4.8\cdot10^{7}$ & $6.7\cdot10^{6}$ & $4.7\cdot10^{7}$ \\
$C_{e} [J/K]$      & $8.2\cdot10^{8}$ & $2.0\cdot10^{9}$ & $1.4\cdot10^{9}$ \\
$A_{in} [m^2]$     & 6.6   & 0.3 & 1.2 \\
$A_{e} [m^2]$      & 7.3  & 41.6 & 19.6 \\
$f_{h}$   & 0.87  & 0.07 & 0.51\\
\hline
\textbf{Training RMSE [W]}   & 361  & 61 & 79 \\
\textbf{Validation RMSE [W]} & 610  & 143 & 103\\
\textbf{Testing RMSE [W]}    & 527  & 270 & 151\\
\hline
\end{tabular}
\end{table}

\subsection{DISTRIBUTION NETWORK SYSTEM MODEL}

\begin{figure}
    \centering
    \includegraphics[width=\columnwidth]{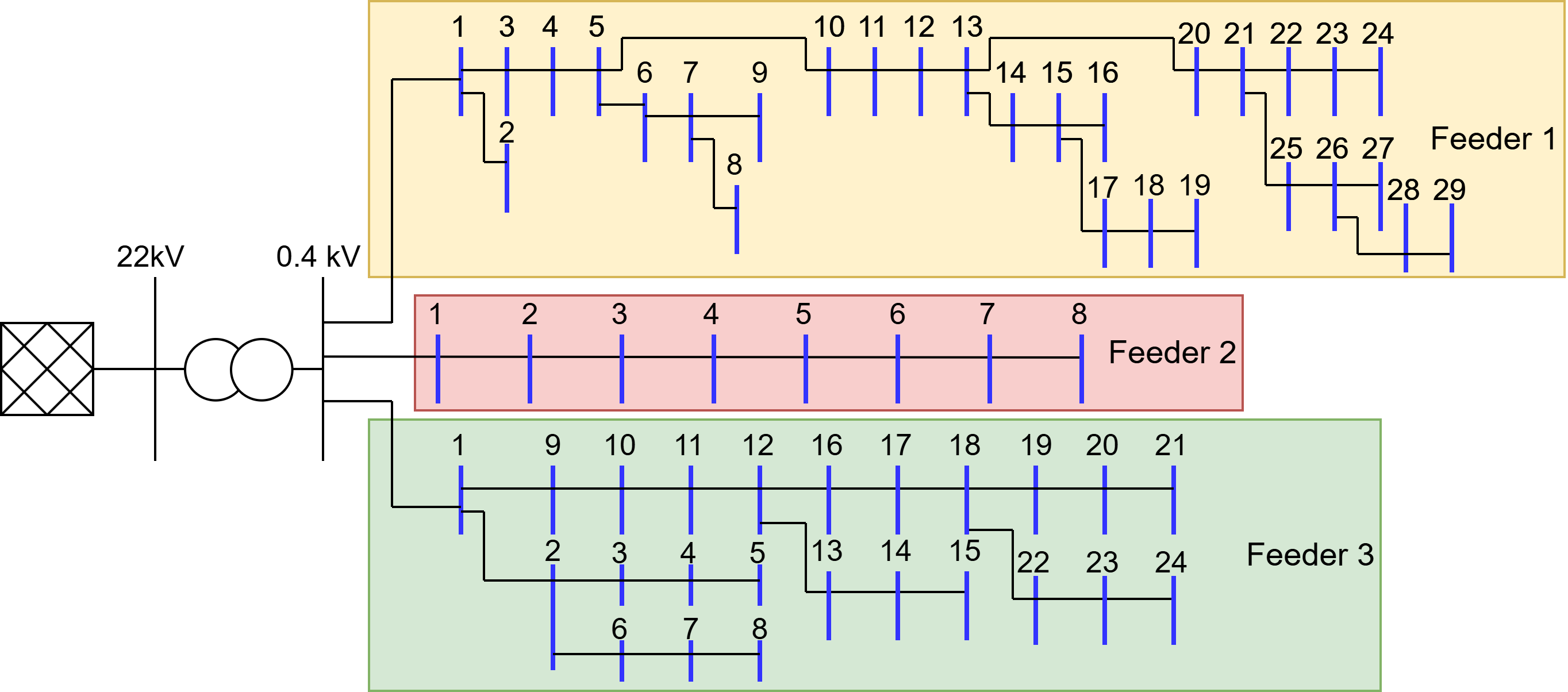}
    \caption{Single-line diagram of the suburban low-voltage distribution network, comprising three feeders connected to a 22/0.4 kV transformer.}
    \label{fig:single-line-diag}
\end{figure}

The analysis was conducted on a realistic LV distribution network provided in \cite{panagi_2025_17369365}. The network consists of a substation equipped with an 11/0.4~kV, 315~kVA delta–wye ($\Delta$–Y) transformer. Three radial feeders extend from the substation, comprising a total of 61 nodes, as depicted in Fig.~\ref{fig:single-line-diag}. Each feeder supplies approximately 30 residential customers.

\section{RESULTS AND DISCUSSIONS}
\label{sec:Results_Discussions}

The SOCP optimization problem was solved using the GUROBI solver, while the IPOPT solver was adopted for the non-linear comparison problem. The implementation was carried out on a laptop equipped with an M4 Pro Max chip, featuring a 12-core CPU and 36 GB of unified memory. For the scheduling operation, a three-day optimization with a $H=30$ minute time interval was used. The first day served as a soft start, the second day as the exact operation study, and the third day was incorporated into the second day's optimization to account for next-day requirements.

\subsection{SCHEDULE OPERATION}

\begin{figure}
    \centering
    \includegraphics[width=1\linewidth]{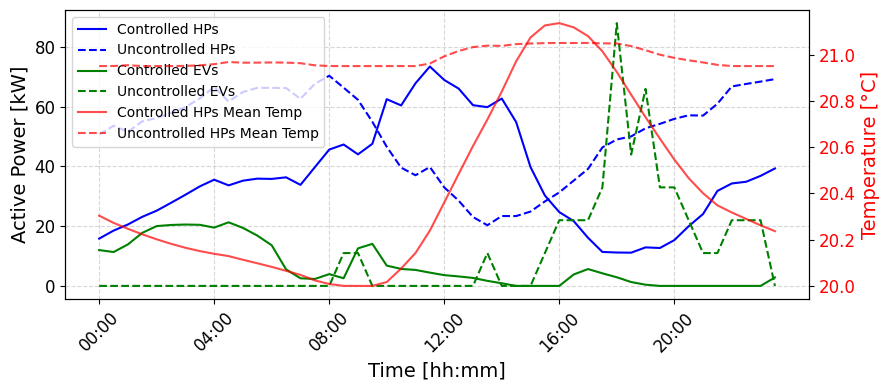}
    \caption{Time-series comparison of active power consumption and indoor temperature for controlled and uncontrolled operation of HPs and EVs.}
    \label{fig:compar_hp_ev_operation}
\end{figure}

Figure~\ref{fig:compar_hp_ev_operation} illustrates the daily power consumption and indoor temperature. Controlled HPs (solid blue line) dynamically adjust demand to follow PV production, increasing operation around noon when solar availability peaks and reducing power during evening hours (16:00–19:00). The indoor temperature remains within comfort limits by exploiting building thermal inertia. In contrast, uncontrolled HPs (dashed blue line) exhibit a relatively flat profile, operating in the opposite direction of PV output and showing limited responsiveness to network conditions.

For EVs, the controlled scenario (solid green line) distributes charging more evenly, with slight emphasis in the early morning to exploit early solar generation and prepare for departures. This strategy avoids the sharp evening peak observed in the uncontrolled case (dashed green line), where vehicles are charged immediately upon arrival, causing load synchronization and increased network stress. These results highlight the benefits of coordinated charging for both peak shaving and PV utilization.

Table~\ref{tab:control_testcases} summarizes the average indoor temperature and total daily energy consumption of HPs and EVs under controllable and uncontrollable operation. Although the average indoor temperature differs only slightly, total energy consumption drops notably (1029~kWh vs. 1486~kWh) in the controllable case. This demonstrates that coordinated control preserves user comfort while reducing energy use for the studied day. This reduction is not associated with a decrease in cumulative energy demand for EV users as they distribute their charging more uniformly over time; consequently, the total energy consumption over an extended horizon remains identical to that of uncontrolled charging. In contrast, heat pumps exhibit a genuine reduction in energy consumption, as coordinated control enables a more effective exploitation of the building’s thermal inertia, leading to smoother intra-day operation compared to conventional control.

Overall, intelligent scheduling preserves user comfort while improving the temporal alignment between demand and local renewable generation, thereby enhancing both energy efficiency and grid operation.

\begin{table}
\caption{Average temperature and energy use under controllable and uncontrollable operation}
\label{tab:control_testcases}
\centering
\begin{tabular}{c|c|c|c}
\hline
\textbf{Type} &
\shortstack{\textbf{Avg Temp} \\ \textbf{(°C)}} &
\shortstack{\textbf{HP Energy} \\ \textbf{(kWh)}}  & 
\shortstack{\textbf{EV Energy} \\ \textbf{(kWh)}} \\
\hline
Controllable   & 20.4 & 866 & 163  \\ \hline
Uncontrollable & 21   & 1227 & 259  \\
\hline
\end{tabular}
\end{table}

\subsection{COMPUTATIONAL COMPLEXITY AND SCALABILITY ASSESSMENT OF OPF FORMULATIONS}

To assess the computational burden and practical scalability of different OPF formulations, a benchmark analysis is conducted across three representative power flow models: the classic Bus Injection Model (BIM), the angle-eliminated DistFlow formulation (Section~\ref{sec:distflow}), and its convex relaxation via SOCP (Section~\ref{sec:conic_relaxation}). The test case evaluates how runtime and solver performance vary with the penetration of EVs, HPs, and PVs in a realistic network.

The results in Table \ref{tab:opf_comparison} reveal a clear trend in the computational burden of the three OPF formulations as the penetration of DERs (EVs, HPs, and PVs) increases. The classic BIM consistently exhibits the highest computational time, with a steep growth from 1.3s at 0\% penetration to 887.7 s at 100\% penetration. This growth rate indicates that BIM scales poorly under high DER penetration, making it less practical for large-scale, real-time applications. The angle-eliminated DistFlow formulation exhibits significantly lower absolute computation times compared to the BIM; however, its scalability trend remains similar, with runtime increasing from 0.46 s to 367.3 s across the same range. The convex DistFlow formulation achieves the lowest computational times for most penetration levels, starting at 0.37 s and rising modestly to 0.87 s for 100\% penetration. 

Compared to BIM and DistFlow, the convex relaxation consistently yields lower objective function values, indicating an improvement in loss minimization performance. From a computational perspective, the convex DistFlow formulation delivers the most significant computational gains, maintaining sub-second runtimes even at the highest penetration levels. This combination of rapid solution time, scalability, and competitive loss minimization performance makes the convex formulation particularly well-suited for operational contexts requiring frequent OPF computations.

\setlength{\tabcolsep}{3pt} 
\begin{table}[!ht]
\caption{Runtime [s] and objective value for OPF formulations (BIM, DistFlow, Convex DistFlow) under increasing DER penetration.}
\label{tab:opf_comparison}
\centering
\begin{tabular}{c|cc|cc|cc}
\hline
\multirow{2}{*}{\textbf{\shortstack{DER\\Penetration}}}
& \multicolumn{2}{c|}{\textbf{BIM}}
& \multicolumn{2}{c|}{\textbf{DistFlow}}
& \multicolumn{2}{c}{\textbf{Convex DistFlow}} \\
\cline{2-7}
& \textbf{Runtime} & \textbf{Obj.}
& \textbf{Runtime} & \textbf{Obj.}
& \textbf{Runtime} & \textbf{Obj.} \\
\hline
0\%   & 1.3   & 0.2624 & 0.46  & 0.2624 & 0.37  & 0.2624 \\ \hline
10\%  & 52.0  & 0.2402 & 36.5  & 0.2503 & 0.42  & 0.2397 \\ \hline
20\%  & 93.1  & 0.2378 & 66.8  & 0.2379 & 0.44  & 0.2328 \\ \hline
40\%  & 228.5 & 0.2355 & 126.5 & 0.2380 & 0.51  & 0.2020 \\ \hline
60\%  & 368.0 & 0.2954 & 174.6 & 0.2863 & 0.64  & 0.2465 \\ \hline
80\%  & 537.8 & 0.3630 & 237.2 & 0.3623 & 0.69  & 0.2731 \\ \hline
100\% & 887.7 & 0.4444 & 367.3 & 0.4473 & 0.87  & 0.3089 \\ \hline
\end{tabular}
\end{table}

\subsection{CONVEXIFICATION EXACTNESS ANALYSIS}

To evaluate the exactness of the SOCP relaxation of the OPF problem, the sufficient conditions provided in Corollary 6 of \cite{li2012exact} are considered. For convenience, these conditions are restated in Appendix \ref{app:convexification_terms}.

Our analysis revealed that none of the sufficient conditions for exactness are fully satisfied in the studied realistic network. Conditions 3, 4, 5, and 6 are violated due to line characteristics and topology. In particular, the presence of service lines with very low reactance ($x$) causes significant deviations in the $r/x$ ratio across consecutive lines, making it impossible to satisfy the cumulative or neighbor-based ratio comparisons required by Conditions 3 and 4. Moreover, Condition 6 fails due to large variations in $r/x$ ratios between neighboring lines. Similarly, Condition 3 is not satisfied because underground lines immediately after the transformer, followed by overhead lines further along the feeder, disrupt the required decreasing trend in $r/x$. Although Condition 2 is topologically satisfied, it prohibits any reverse power flow, a constraint invalid under operating scenarios with maximum distributed generation and minimum load. Finally, Condition~1 cannot be guaranteed in practice due to limited controllability and demand-response availability across all buses, preventing enforcement of strictly positive nominal active and reactive powers at every node.

Nevertheless, the studied case exhibited exactness, verified after solving the problem. This may result from other sufficient conditions, not explicitly stated here, being satisfied, or from some of the conservative conditions actually holding under a more accurate network representation. In general, for LV networks that do not explicitly model service lines and lack capacitive reactive power control, Conditions 4 and 5 can be more easily satisfied. 

Another significant point well established in the literature is that the exactness of SOCP relaxations of the DistFlow is really sensitive to the voltage-rise phenomena associated with high distributed generation and low loading levels. In such cases, the additional degrees of freedom introduced by the relaxation may allow non-physical operating points, for instance, through artificial inflation of line currents and losses, which can indirectly affect voltage profiles. This limitation has been extensively discussed in prior studies. 

In the present work, the analyzed case study is characterized by a relatively high baseline demand, further increased by the electrification of heating and transportation via heat pumps and electric vehicles. Combined with the specific cable characteristics of the network, these conditions prevent voltage rise events. As a result, the operating regime considered in this study does not activate the known pathological behaviors of the SOCP relaxation. Nevertheless, readers should be aware of applications where overvoltage conditions are expected.
When obtaining an exact solution is critical, iterative methodologies, some based on the same branch flow model, can be applied instead of the single-run OPF formulation proposed in this work. For example,~\cite{andrianesis2021optimal} presents a fast-solving approach that provides an efficient alternative under such scenarios.

\subsection{DISTRIBUTION NETWORK OPERATION IMPROVEMENTS}

\subsubsection{LOSSES}

\begin{figure}
    \centering
    \includegraphics[width=1\linewidth]{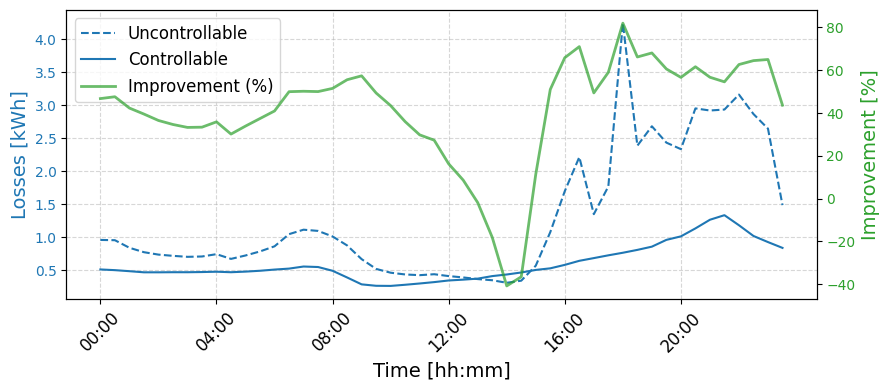}
    \caption{Time-series comparison of total losses (transformer + line) under controllable and uncontrollable operation. The green curve shows the percentage improvement achieved via coordinated scheduling.}
    \label{fig:combined_losses}
\end{figure}

The combined transformer and line losses over the day are shown in Fig.~\ref{fig:combined_losses}. The proposed controllable operation reduces daily losses from 1.29\% to 0.59\%, a 54\% improvement in overall efficiency. Throughout the day, the coordinated strategy maintains lower losses than the uncontrolled case. Between 13:00 and 15:00, a brief increase appears due to higher HP operation, while in the uncontrolled most HPs operate lower during this period due to the reduced HP thermal demand and limited use of thermal inertia, resulting in a flatter profile. The largest gains occur during the evening peak (16:00–20:00), when coordination suppresses simultaneous EV charging and HP usage, minimizing transformer loading and line currents. The green trace in Fig.~\ref{fig:combined_losses} illustrates the percentage improvement over time.

\subsubsection{VOLTAGE LEVELS}

\begin{figure}
    \centering
    \includegraphics[width=1\linewidth]{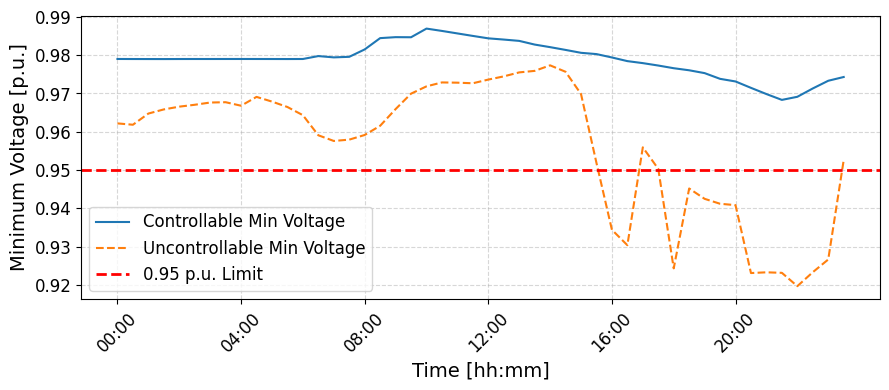}
    \caption{Minimum voltage profile throughout the day for controllable and uncontrollable operation.}
    \label{fig:voltage_levels}
\end{figure}

Figure~\ref{fig:voltage_levels} shows the minimum voltage trajectory over the day for both control strategies. The controllable case maintains voltage levels closer to nominal values. This improvement results from the DERs' reactive power contribution, which offsets reactive current demand, thereby reducing total current and associated voltage drops. Additionally, the improvement reflects the effectiveness of coordinated active power operation compared to the uncontrolled case. The number of voltage violations is also reduced relative to the uncontrolled scenario. The lowest voltages, reaching 0.92~p.u., occur during the evening peak (17:00–21:00) in the absence of coordinated control, indicating severe violations of grid operational limits.

\subsubsection{TRANSFORMER LOADING AND AGING}

\begin{figure}
    \centering
    \includegraphics[width=1\linewidth]{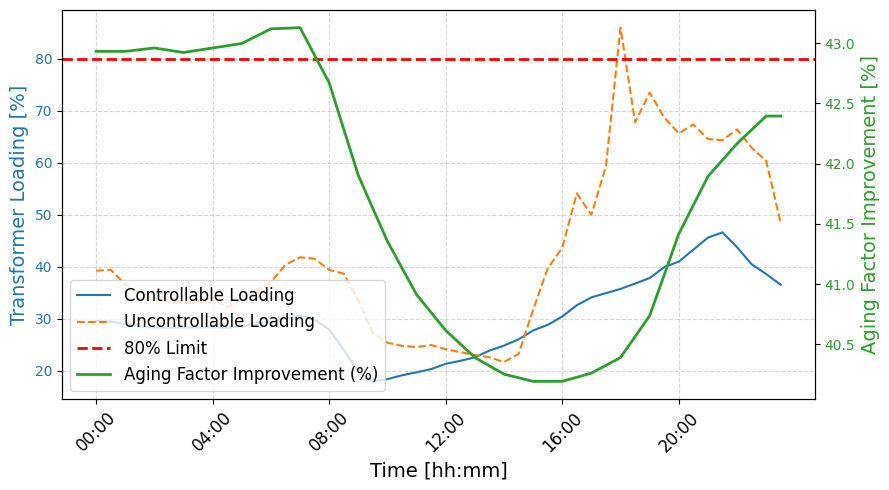}
    \caption{Transformer loading over the day for controllable and uncontrollable operation.}
    \label{fig:transformer_loading}
\end{figure}

The transformer loading and aging improvement factor is shown in Fig.~\ref{fig:transformer_loading}. Transformer loading is significantly reduced under the controllable strategy. While uncontrolled operation exceeds the typical 80\% loading threshold during peak hours, the coordinated case maintains loading below this limit throughout the day. This reduction is critical for ensuring transformer longevity and safe operation. The evening peak is particularly pronounced in the uncontrolled case, where simultaneous EV and HP operation causes sharp loading surges.

The transformer aging metric, as described in Section~\ref{sec:transformer_aging}, indicates an average daily improvement of 41\%. This implies that the lifetime of transformer windings increases compared to the reference case when controllable operation is applied. Although transformer loading under the controllable strategy may temporarily exceed the corresponding values in the uncontrolled scenario, the overall thermal stress remains significantly lower. This is due to the exponential dependence of insulation aging on the winding hottest-spot temperature, which incorporates thermal inertia and previous operating conditions. Therefore, short and well-managed load increases do not necessarily lead to higher degradation.

\subsubsection{LINE LOADING}

\begin{figure}
    \centering
    \includegraphics[width=1\linewidth]{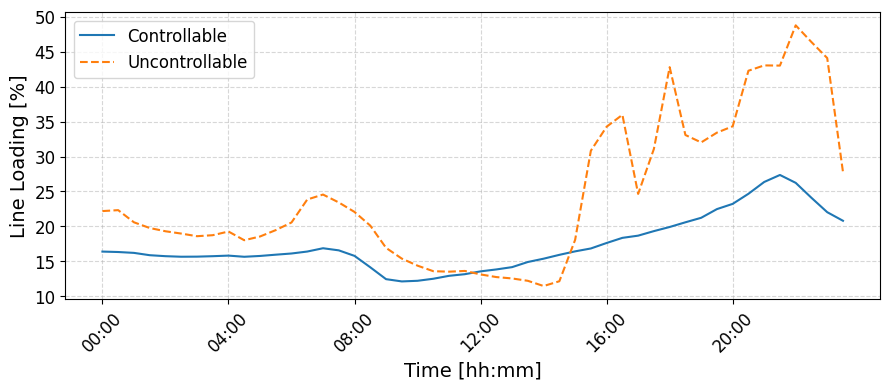}
    \caption{Maximum line loading comparison for controllable and uncontrollable operation.}
    \label{fig:line_loading}
\end{figure}

The Fig.~\ref{fig:line_loading} shows the maximum line loading profiles for both scenarios. While no thermal violations occur due to the relatively high ampacity margins of the studied network, the controllable operation consistently yields lower loading, particularly during the evening peak. This reduction demonstrates the capability of coordinated DERs to alleviate grid congestion. In more constrained systems, such reductions would be critical for avoiding line overloads and ensuring operational reliability.

\section{CONCLUSIONS}
\label{sec:conclusions}

This paper presented a comprehensive modeling and optimization framework for the coordinated operation of EVs, HPs, and PV systems in active distribution networks. By integrating a calibrated 3R2C grey-box building thermal model into a network-constrained OPF, the proposed method establishes a direct link between thermal comfort requirements and electrical flexibility from HPs. A comparative assessment of OPF formulations demonstrated that the convexified DistFlow model offers substantial computational advantages over non-convex formulations while maintaining competitive optimality performance. The SOCP relaxation analysis under realistic LV feeder conditions revealed that theoretical exactness guarantees are often not satisfied in practice, though exactness was observed in the studied case. Simulation results on a real Cypriot LV network showed that the coordinated strategy significantly improves grid performance by reducing transformer aging by up to 41\%, cutting daily losses by 54\%, lowering voltage violations, and alleviating line loading. These findings confirm that integrated DER scheduling is a key enabler for enhancing both operational reliability and energy efficiency in future electrified distribution systems. 

Future work will also explore the integration of Model Predictive Control (MPC) strategies to enable real-time, receding-horizon optimization of EV and HP operation. This extension will allow the framework to adapt dynamically to forecast deviations, enhance robustness under uncertainty, and bridge the gap between day-ahead scheduling and real-time control.

\bibliographystyle{IEEEtran}
\bibliography{biblio.bib} 

\appendix
For reference, this appendix summarizes the sufficient conditions derived in Corollary 6 of \cite{li2012exact}, which establish when the SOCP relaxation of the OPF problem is exact. These conditions ensure that, if any of them is satisfied, the solution obtained from the relaxed formulation coincides with the optimal solution of the original non-convex OPF. The conditions are listed below:
\begin{itemize}
    \item[(1)] For each \( j \in \mathcal{N} \setminus \{0\} \), \( \overline{p_j} \leq 0 \), \( \overline{q_j} \leq 0 \).
    \item[(2)] For all \( (i,j) \in \mathcal{L} \), \( \frac{r_{i,j}}{x_{i,j}} \geq \frac{R_i}{X_i} \), with \( \overline{p_j} \leq 0 \).
    \item[(3)] For all \( (i,j), (j,k) \in \mathcal{L} \), \( \frac{r_{i,j}}{x_{i,j}} \geq \frac{r_{j,k}}{x_{j,k}} \), with \( \overline{p_j} \leq 0 \).
    \item[(4)] For all \( (i,j) \in \mathcal{L} \), \( \frac{x_{i,j}}{r_{i,j}} \geq \frac{X_i}{R_i} \), with \( \overline{q_j} \leq 0 \).
    \item[(5)] For all \( (i,j), (j,k) \in \mathcal{L} \), \( \frac{x_{j,k}}{r_{j,k}} \geq \frac{x_{i,j}}{r_{i,j}} \), with \( \overline{q_j} \leq 0 \).
    \item[(6)] For all \( (i,j), (j,k) \in \mathcal{L} \), \( \frac{r_{j,k}}{x_{j,k}} = \frac{r_{i,j}}{x_{i,j}} \).
\end{itemize}
where $R_i$ and $X_i$ define the cumulative resistance and reactance from the root node to bus $i$.
\label{app:convexification_terms}

\end{document}